\documentclass[prd,showpacs,preprintnumbers,twocolumn]{revtex4}
\usepackage{amsmath}
\usepackage{graphicx}
\usepackage{amsfonts}
\usepackage{amssymb}

\begin{document}

\title{Nonlinear self-interaction of plane gravitational waves} 

\author{M.\ Servin$^1$, M.\ Marklund$^2$, G.\ Brodin$^1$, J.\ T.\
  Mendon\c{c}a$^3$ and V.\ Cardoso$^4$}

\affiliation{$1$~Department of Plasma Physics, Ume{\aa} University,
  SE--901 86 Ume{\aa}, Sweden}

\affiliation{$2$~Department of Electromagnetics, Chalmers University
  of Technology, 
  SE--412 96 G\"oteborg, Sweden}

\affiliation{$3$~GoLP, Instituto Superior T\'ecnico, 1049--001 Lisboa,
  Portugal}

\affiliation{$4$~CENTRA, Departemento de F\'{\i}sica, Instituto
  Superior 
  T\'ecnico, Av.\ Rovisco Pais 1, 1049--001 Lisboa,
  Portugal}

\begin{abstract}
Recently Mendon\c{c}a and Cardoso [Phys.\ Rev.\ D \textbf{66},
104009 (2002)] considered nonlinear gravitational wave
packets propagating in flat space-time. They concluded that the
evolution 
equation---to third order in amplitude---takes a similar form to what
arises 
in nonlinear optics. Based on this equation, the authors found that
nonlinear 
gravitational waves exhibit self-phase modulation and high harmonic
generation 
leading to frequency up-shifting and spectral energy dilution of the  
gravitational wave energy. In this Brief Report we point out the fact---a
possibility that seems to have been overlooked by Mendon\c{c}a and
Cardoso---that the nonlinear terms in the evolution equation cancels
and, hence, that there is no amplitude evolution of the pulse. Finally we
discuss scenarios where these nonlinearities may play a role.
\end{abstract}

\pacs{04.30.Nk, 42.65.-k, 95.30.Sf}

\maketitle

As discussed by Mendon\c{c}a and Cardoso in \cite{1} the Einstein
field equations are 
highly nonlinear and it is a most interesting question whether gravitational
waves share any similarities with other types of more well-known nonlinear
waves, for instance those occurring in nonlinear optics. Assuming the presence
of a number of wave perturbations
\begin{equation}
  h_{ab}(n)=\epsilon_{ab}(n)a(n)\exp[iq_{c}(n)x^{c}] + \mathrm{c.c.}
  \label{pert} , 
\end{equation}
it was---based on the structure of the vacuum gravitational field
equations 
---argued by Mendon\c{c}a and Cardoso that the wave amplitude evolution 
equations for four-wave mixing 
processes take the form
\begin{equation}
iq_{a}(1)\frac{\partial}{\partial
  x_{a}}a(1)=w(1)a^{\ast}(2)a(3)a(4)\exp 
[i\Delta q_{b}x^{b}] , \label{Mendonca}%
\end{equation}
where $(n)$ labels the wave perturbations, $\epsilon_{ab}(n)$ is the
unit 
polarization tensor for wave $(n)$, $a(n)$ is the slowly varying
amplitude 
($\ast$ stands for complex conjugate), $q_{c}(n)$ is the wave
four-vector, 
c.c.\ denotes the complex conjugate of the preceding term, $\Delta
q_{b}\equiv q_{b}(3)+q_{b}(4)-q_{b}(1)-q_{b}(2)$ and the coupling
coefficient 
$w(1)$ is determined by the Ricci tensor cubic in the wave amplitudes
(not 
explicitly computed in \cite{1}). The background space-time was
assumed flat. The 
authors then considered the implications of Eq.\ (\ref{Mendonca}),
which indeed 
is similar to what arises in nonlinear optics. The analysis was
restricted to the case when all present waves propagate in the same
direction.\ In particular they focused on (i) \emph{self-phase
  modulation} 
(when the four waves have identical polarization and wave vector)
resulting in 
that a gravitational wave pulse will be continuously frequency
up-shifted as 
it propagates through space and (ii) \emph{harmonic cascading} through
which 
gravitational wave energy is spectrally diluted.

In this Brief Report we present the results of detailed calculations of the
amplitude evolution equation for plane gravitational waves propagating
along 
$x^{3}\equiv z$. The perturbed metric $g_{ab}$ is
\begin{equation}
g_{ab}=\eta_{ab}+h_{ab}^{(1)}+h_{ab}^{(2)}+h_{ab}^{(3)} ,
\end{equation}
where $\eta_{ab} = \text{diag}(-1,1,1,1)$, $h_{ab}^{(2)}$ and
$h_{ab}^{(3)}$ 
are the second and third order nonlinear response to the linear wave
perturbation $h_{ab}^{(1)}$. 
As it turns out, there 
are no third order response terms that couple back to the evolution
equation 
of $h_{ab}^{(1)}$, therefore $h_{ab}^{(3)}$ is omitted from here
on. The vacuum Einstein field equations 
reads
\begin{equation}
R_{ab}^{(1)}+R_{ab}^{(2)}+R_{ab}^{(3)}=0\label{Ricci} ,
\end{equation}
where the Ricci tensor has been divided into $R_{ab}^{(1)}$ (linear in
$h_{ab}^{(1)}$), $R_{ab}^{(2)}$ (quadratic in $h_{ab}^{(1)}$ and
linear in 
$h_{ab}^{(2)}$) and $R_{ab}^{(3)}$ (cubic in $h_{ab}^{(1)}$ and with
terms of 
order $h_{ab}^{(1)}h_{ab}^{(2)}$). Having wave perturbations of the
type given by
Eq.\ (\ref{pert}) in mind we note that in $R_{ab}^{(3)}$, but not in
$R_{ab}^{(2)}$, there are terms that are resonant with
$R_{ab}^{(1)}$. Thus, we identify
\begin{equation}
  R_{ab}^{(1)}=-R_{ab}^{(3)}\label{R13} 
\end{equation}
with the amplitude evolution equation, where it should be understood
that only 
the part of $R_{ab}^{(3)}$ resonant with $R_{ab}^{(1)}$ should be
taken into 
account. The second order nonlinear response terms, $h_{ab}^{(2)}$,
are 
related to the $h_{ab}^{(1)}$ by
\begin{equation}
  R_{ab}^{(2)}=0\label{R2} .
\end{equation}
We are free to choose a particular gauge for $h_{ab}^{(1)}$ and we
choose the 
TT gauge such that the only nonzero components are
$h_{11}^{(1)}=-h_{22}%
^{(1)}\equiv h$ \cite{2}. With this choice $R_{ab}^{(2)}$ reads
\begin{subequations}\label{R2A}
\begin{align}
R_{00}^{(2)} &  =\frac{1}{2}\left[  -\lambda-\partial_{t}^{2}H\right] 
+\frac{1}{2}\left(  \partial_{t}h\right)
^{2}+h\partial_{t}^{2}h , \\ 
R_{03}^{(2)} &
=-\frac{1}{2}\partial_{z}\partial_{t}H+\frac{1}{2}\left( 
\partial_{t}h\right)  \left(  \partial_{z}h\right)
+h\partial_{z}\partial 
_{t}h , \\
R_{33}^{(2)} &  =\frac{1}{2}\left[  \lambda-\partial_{z}^{2}H\right] 
+\frac{1}{2}\left(  \partial_{z}h\right)
^{2}+h\partial_{z}^{2}h , 
\end{align}
\end{subequations}
and
\begin{subequations}\label{R2B}
\begin{align}
R_{0\alpha}^{(2)} &  =\frac{1}{2}\left[
  -\partial_{z}^{2}h_{0\alpha}^{(2)}+\partial 
_{z}\partial_{t}h_{3\alpha}^{(2)}\right] , \\
R_{3\alpha}^{(2)} &  =\frac{1}{2}\left[
  \partial_{t}^{2}h_{3\alpha}^{(2)}-\partial 
_{z}\partial_{t}h_{0\alpha}^{(2)}\right] , \\
R_{\alpha\beta}^{(2)} &  =\frac{1}{2}\left[
  \partial_{t}^{2}-\partial_{z}^{2}\right] 
h_{\alpha\beta}^{(2)}-\frac{1}{2}\left[  \left(  \partial_{t}h\right)
  ^{2} - \left( 
\partial_{z}h\right)  ^{2}\right]  \delta_{\alpha\beta} , 
\end{align}
\end{subequations}
where $\alpha,\beta=1,2$ and
\begin{align}
H &  \equiv h_{11}^{(2)}+h_{22}^{(2)} ,\\
\lambda &
\equiv\partial_{z}^{2}h_{00}^{(2)}-2\partial_{z}\partial_{t}%
h_{03}^{(2)}+\partial_{t}^{2}h_{33}^{(2)} .
\end{align}
When computing $R_{ab}^{(2)}$ and $R_{ab}^{(3)}$ one should make use
of the 
approximation $\partial_{z}\approx-\partial_{t}$ \cite{dz}.
Applying this to 
$R_{ab}^{(2)}$ it follows that all the components of $h_{ab}^{(2)}$
may be set 
to zero except $h_{11}^{(2)}$ and $h_{22}^{(2)}$ (the final result
does not depend 
on whether this choice, corresponding to a freedom in gauge, is made
or not). 
Making this choice Eq.\ (\ref{R2}) becomes
\begin{equation}
\partial_{t}^{2}H=-\left(  \partial_{t}h\right)^{2}+2\partial_{t}^{2}%
(h)^{2}\label{H} .
\end{equation}
The nonzero components of $R_{ab}^{(3)}$ are
\begin{subequations}
\begin{align}
R_{11}^{(3)} &  =\frac{1}{4}(\partial_{z}h)(\partial_{z}H)-\frac{1}%
{4}(\partial_{t}h)(\partial_{t}H) , \label{R3_11}\\
R_{22}^{(3)} &  =-\frac{1}{4}(\partial_{z}h)(\partial_{z}H)+\frac{1}%
{4}(\partial_{t}h)(\partial_{t}H) , \label{R3_22}%
\end{align}
\end{subequations}
and with the approximation $\partial_{z}\approx-\partial_{t}$ we see that
$R_{ab}^{(3)}=0$. The amplitude evolution equation (\ref{R13}) thus reduces
to
\begin{equation}
(\partial_{z}^{2}-\partial_{t}^{2})h=0\label{Dh} ,
\end{equation}
and we conclude that---up to third order in amplitude---plane gravitational
waves propagate with no change in the amplitude modulation, i.e., $w(1)$ in 
Eq.\ (\ref{Mendonca}) is zero. To some extent this may seem
counterintuitive, since a gravitational wave-packet propagating in flat
space-time transports energy and momentum as described by the pseudo
energy-momentum tensor which is of order $h^2$. This produces a local
curvature of space-time of the same order, and one could expect a
coupling to the original wave perturbation, leading to cubic
nonlinearities in the amplitude evolution equation. However, this
picture is only 
partially true. The pseudo energy-momentum tensor does produce curvature
quadratic in the amplitude $h$ 
(essentially described by Eq.\ (\ref{H})), and this
couples back to the original perturbation leading to cubic nonlinearities
in the Riemann tensor. But as follows from Eq.\ (\ref{R3_11}) and 
Eq.\ (\ref{R3_22}) these terms cancels when computing the Ricci tensor
and thus does not enter the amplitude evolution equation.

It is instructive to compare this result to existing exact plane
wave solutions of the vacuum Einstein field. One particular exact plane wave
solution is given by the metric (see, e.g., Ref.\ \cite{3})
\begin{equation}
  ds^{2}=-dt^{2}+a(\xi)^{2}dx^{2}+b(\xi)^{2}dy^{2}+dz^{2} ,
\end{equation}
where $a$ and $b$ satisfy the equation
\begin{equation}
  b\partial_{\xi}^{2}a+a\partial_{\xi}^{2}b=0\label{PPR} ,
\end{equation}
where $\xi=z-t$. Focusing on weak waves, we take $a=(1+h+\frac{1}{2}%
H)^{\frac{1}{2}}$ and $b=(1-h+\frac{1}{2}H)^{\frac{1}{2}}$ where $1\gg
h$ and
$H\sim h^{2}$. Expanding Eq.\ (\ref{PPR}) up to third order in amplitude one
find precisely the relation (\ref{H}) between $H$ and $h$ and that the third
order terms cancels identically. The fact that the metric components of the
gravitational wave perturbations in Eq.\ (\ref{pert}) (with the gauge choices
made above) coincide with those of the exact wave solution when expanded in
powers of amplitude suggests that Eq.\ (\ref{Dh}) will remain even if one
relaxes the approximations made in this Brief Report.

It should be pointed out, however, that gravitational four-wave mixing
processes 
are not ruled out in general. Firstly, we have here only considered
parallel 
propagation, and it is well-known that anti-parallel gravitational
waves can  
interact (see e.g.\ Ref.\ \cite{4} and references therein). 
Secondly, the presence of background 
curvature and matter modifies the properties of gravitational
waves, e.g., giving rise to wave dispersion and new types of nonlinearities.
To what extent this increases the possibilities of nonlinear
self interaction 
is an open question. The equations presented here provides a starting
point for future investigations of these possibilities.


\begin{thebibliography}{99}
\bibitem{2} Due to the assumption of slowly varying amplitudes
  $h_{11}^{(1)}$ and $h_{22}^{(1)}$ are not the only nonzero
  components. The other components are, however, of higher order and
  these terms can therefore be absorbed in $h_{ab}^{(3)}$.
\bibitem{dz} One might think that the use of $\partial_{z}\approx-\partial_{t}$
 is too restrictive. However, improving this approximation produces necessarily 
terms in the amplitude evolution equation that are of higher order than cubic.
\end{thebibliography}
\end{document}